\documentclass[preprint]{acmart}
\AtBeginDocument{%
  \providecommand\BibTeX{{%
    \normalfont B\kern-0.5em{\scshape i\kern-0.25em b}\kern-0.8em\TeX}}}

\setcopyright{rightsretained}
\acmDOI{http://dx.doi.org/10.1145/3434642}

\acmJournal{CACM}
\acmVolume{-}
\acmNumber{-}

\usepackage{xcolor}

\begin{document}

\title{The Future is Big Graphs! A Community View on 
Graph Processing Systems}

\author{Sherif Sakr}
\authornote{Unfortunately, the first author passed away during the period following the event and the completion of this article. Thus, this article is published in memoriam.}
\email{sherif.sakr@ut.ee}
\affiliation{%
  \institution{University of Tartu}
  \country{Estonia}
}

\author{Angela Bonifati}
\affiliation{%
  \institution{Lyon 1 University \& Liris CNRS}
  \country{France}}
\email{angela.bonifati@univ-lyon1.fr}

\author{Hannes Voigt}
\affiliation{%
  \institution{Neo4j}
  \country{Germany}
}
\email{hannes.voigt@neo4j.com}

\author{Alexandru Iosup}
\affiliation{%
 \institution{Vrije Universiteit Amsterdam \& Delft University of Technology}
 \country{the Netherlands}}
\email{A.Iosup@vu.nl}

\author{the computer systems and data management communities}
\email{seminar-19491@dagstuhl.de}
\authornote{
Full author list:
Sherif Sakr (University of Tartu, EE);
Angela Bonifati (Lyon 1 University \& Liris CNRS, FR);
Hannes Voigt (Neo4j, DE);
Alexandru Iosup (Vrije Universiteit Amsterdam \& Delft University of Technology, NL);
Khaled Ammar (BorialisAI, CA);
Renzo Angles (University of Talca – Chile, CL);
Walid Aref (Purdue University – West Lafayette, US);
Marcelo Arenas (PUC \& IMFD – Santiago de Chile, CL); 
Maciej Besta (ETH Zürich, CH);
Peter A. Boncz (CWI – Amsterdam, NL);
Khuzaima Daudjee (University of Waterloo, CA);
Emanuele Della Valle (Polytechnic University of Milan, IT);
Stefania Dumbrava (ENSIIE – Evry, FR);
Olaf Hartig (Linköping University, SE);
Bernhard Haslhofer (AIT – Austrian Institute of Technology – Wien, AT);
Tim Hegeman (VU University Amsterdam, NL);
Jan Hidders (Birkbeck, University of London, GB); 
Katja Hose (Aalborg University, DK);
Adriana Iamnitchi (University of South Florida – Tampa, US);
Vasiliki Kalavri (Boston University, US);
Hugo Kapp (Oracle Labs Switzerland – Zürich, CH);
Wim Martens (Universität Bayreuth, DE);
M. Tamer Özsu (University of Waterloo, CA); 
Eric Peukert (Universität Leipzig, DE); 
Stefan Plantikow (Neo4j – Berlin, DE);
Mohamed Ragab (University of Tartu, EE); 
Matei R. Ripeanu (University of British Columbia – Vancouver, CA); 
Semih Salihoglu (University of Waterloo, CA);
Christian Schulz (Heidelberg University, DE, and Universität Wien, AT); 
Petra Selmer (Neo4j – London, GB); 
Juan F. Sequeda (data.world – Austin, US); 
Joshua Shinavier (Uber Engineering – Palo Alto, US); 
Gábor Szárnyas (Budapest Univ. of Technology and Economics, HU);
Riccardo Tommasini (University of Tartu, EE);
Antonino Tumeo (Pacific Northwest National Lab. – Richland, US); 
Alexandru Uta (VU University Amsterdam, NL); 
Ana Lucia Varbanescu (University of Amsterdam, NL); 
Hsiang-Yun Wu (TU Wien, AT); 
Nikolay Yakovets (TU Eindhoven, NL); 
Da Yan (The University of Alabama – Birmingham, US); 
Eiko Yoneki (University of Cambridge, GB). 
}
\authornote{This article has been accepted for publication by the Communications of the ACM, in November 2020. The full publication should take place in the next few months. For now, we have been given a DOI; if it changes during the publication process, we will update this document to include the proper citation.}


\renewcommand{\shortauthors}{Sakr, Bonifati, Voigt, Iosup, et al.}


\maketitle

\section{Introduction}

Graphs are by nature ‘unifying abstractions’ that can leverage interconnectedness to represent, explore, predict, and explain real- and digital-world phenomena. Although real users and consumers of graph instances and graph workloads understand these abstractions, future problems will require new abstractions and systems. What needs to happen in the next decade for big graph processing to continue to succeed?

We are witnessing an unprecedented growth of interconnected data, which underscores the vital role of graph processing in our society. Instead of one single, exemplary (“killer”) application, we see big graph processing systems underpinning many emerging, but already complex and diverse data management ecosystems, in many areas of societal interest\footnote{As indicated by a user survey [12] and by a systematic literature survey of 18 application domains, including biology, security, logistics and planning, social sciences, chemistry, and finance, see \url{http://arxiv.org/abs/1807.00382}}. To name only a few remarkable examples of late, the importance of this field for practitioners is evidenced  by the large number (over 60,000) of people registered\footnote{Cf. \url{https://app.databox.com/datawall/551f309602080e2b2522f7446a20adb705cabbde8}} to download the Neo4j book “Graph Algorithms”\footnote{\url{ https://www.oreilly.com/library/view/graph-algorithms/9781492047674/}} in just over 1.5 years, and by the enormous interest in the use of graph processing in the Artificial Intelligence and Machine Learning fields.\footnote{Many highly cited articles support this statement, e.g, Hamilton, W., Zhitao, Y., and Leskovec, J. Inductive representation learning on large graphs. NIPS (2017); Perozzi, B., Al-Rfou, R., Skiena S. DeepWalk: Online Learning of Social Representations. \url{https://arxiv.org/pdf/1403.6652.pdf}} Furthermore, the timely Graphs4Covid-19 initiative\footnote{\url{https://neo4j.com/graphs4good/covid-19/}} provides evidence for the importance of big graph analytics in alleviating the global COVID-19 pandemic.

To address the growing presence of graphs, academics, start-ups, but also big tech companies such as Google, Facebook, and Microsoft, have introduced various systems for managing and processing big graphs. Google’s PageRank (late-1990s) showcased the power of web-scale graph processing; and motivated the development of the MapReduce programming model, which was originally used to simplify the construction of the data structures used to handle searches, but has since been used extensively outside of Google to implement algorithms for large-scale graph processing. Motivated by scalability, Google Pregel model of “think like a vertex” (2010) enabled distributed PageRank computation, while the Facebook and Apache Giraph and ecosystem extensions support more elaborate computational models (i.e., task-based and not always distributed) and data models (i.e., diverse, possibly streamed, possibly wide-area data sources) useful for social network data. Simultaneously, an increasing number of use cases revealed RDBMS performance problems on management of highly connected data, motivating various startups and innovative products, such as Neo4j, Sparksee, and the current Amazon Neptune. Microsoft Trinity and later Azure SQL DB provided an early distributed database-oriented approach to big graph management. 

The diversity of models and systems led initially to the fragmentation of the market and a lack of clear direction for the community. Opposing this trend, we see promising efforts to bring together the programming languages, the ecosystem structure, and the performance benchmarks. As we have argued, there is no single most-common (“killer”) application, so the community cannot be brought together around it. 

Co-authored by a representative sample of the community, this article addresses the questions: How do the next-decade big graph processing systems look like from the perspectives of the data management and the large scale systems communities\footnote{The summary of the Dagstuhl seminar: \url{https://www.dagstuhl.de/19491}}? What can we say today about the guiding design principles of these systems in the next 10 years?

\begin{figure}[!t]
  \centering
  \includegraphics[width=\linewidth]{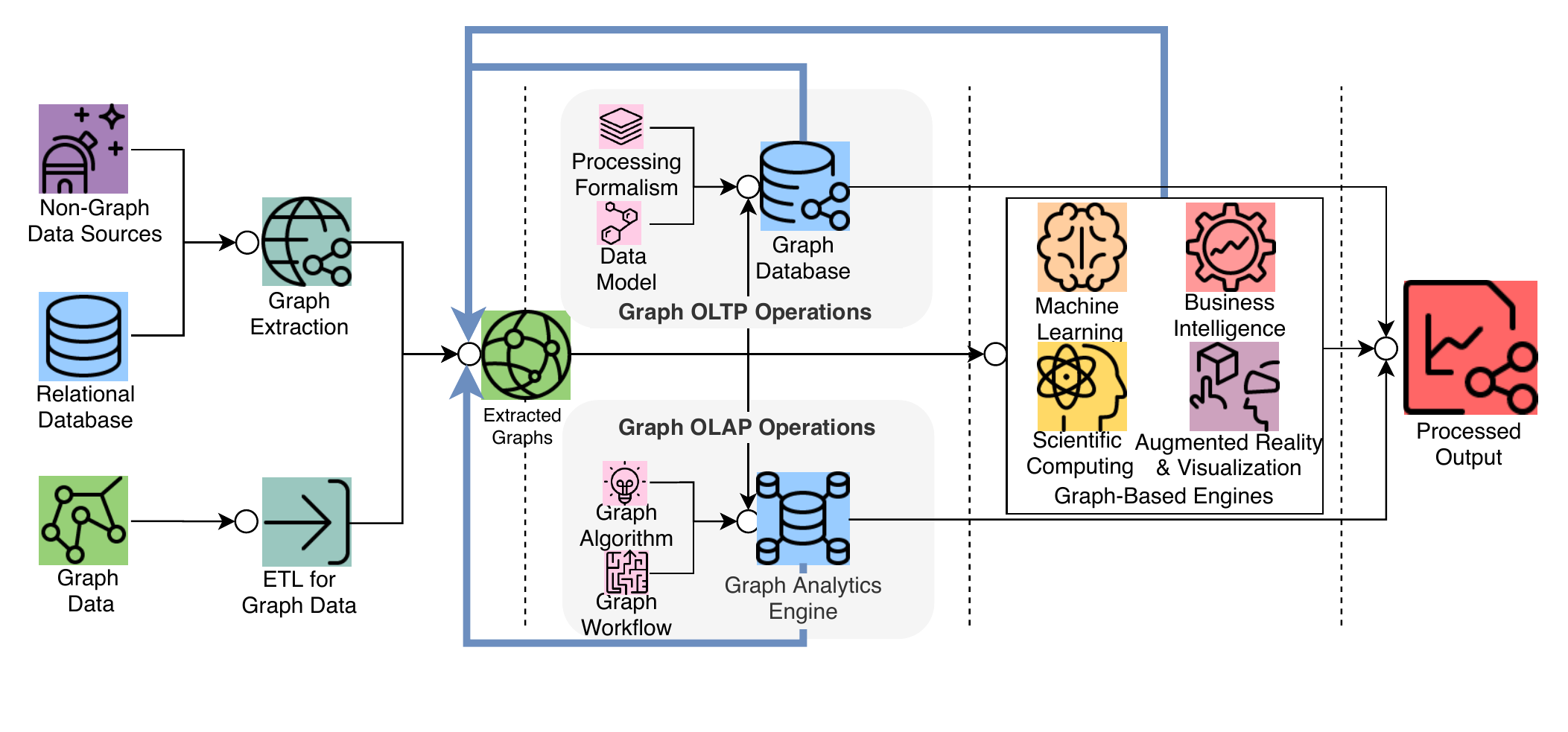}
  \caption{Illustration of a complex data pipeline for graph processing. Data is flowing left to right, from data source to output, though a series of functionally different processing steps. Feedback and loopbacks flow mainly through the blue (highlighted) arrows.}
  \Description{A complex data pipeline for graph processing.}
  \label{fig:graphpipelines}
\end{figure}

Figure~\ref{fig:graphpipelines} outlines the complex pipeline of future big graph processing systems. Data flows in from diverse sources (already graph-modelled as well as non-graph-modelled), is persisted, managed and manipulated with OnLine Transactional Processing operations, such as insertion, deletion, update, filtering, projection, joining, uniting, intersecting, analysed, enriched, and condensed with OnLine Analytical Processing operations, such as grouping, aggregating, slicing, dicing, rollup, and gets disseminated and consumed by machine learning (e.g., ML libraries and processing frameworks), business intelligence (e.g., report generating and planning tools), scientific computing, and visualization and augmented reality applications (for inspection and interaction by the user). Note that this is typically not a purely linear process and hybrid OLTP/OLAP processes can emerge. Considerable complexity stems from (intermediate) results being fed back into early process steps, as indicated by the blue arrows. 

As an example, to study coronaviruses and their impact on the human and animal populations (e.g., the COVID-19 disease), the pipeline depicted in Figure~\ref{fig:graphpipelines} could be purposed for two major kinds of analysis: network-based ‘omics’ and drug-related search, and network-based epidemiology and spread-prevention. For the former, the pipeline could have the following steps: the initial genome sequencing leads to identifying similar diseases, then text (non-graph data) and structured (database) searches help identify genes related to the disease, then a network treatment coupled with various kinds of simulations could reveal various drug targets and valid inhibitors, could lead to effective prioritization of usable drugs and treatments. For the latter, social media and location data, and data from other privacy-sensitive sources, could be combined into social interaction graphs, which could be traversed to establish super-spreaders and super-spreading events related to them, leading to prevention policies and containment actions.

However, the current generation of graph processing technology cannot support such a complex pipeline. For instance, on the COVID-19 knowledge graph,\footnote{\url{https://covidgraph.org/}} even though useful queries can be posed against individual graphs\footnote{\url{https://github.com/covidgraph/documentation/blob/master/helpful-queries.md}} inspecting the papers, patents, genes, and most influential authors related to COVID-19, a full-fledged graph processing pipeline across multiple (graph-)datasets as illustrated in Figure~\ref{fig:graphpipelines} inspecting several data sources raises many challenges for the current graph database technology. In this paper, we formulate these challenges and build our vision for next-generation, big graph processing systems by focusing on three major aspects: abstractions, ecosystem, and performance.

First, we present expected data models and query languages, and inherent relationships among them in lattice of abstractions. We discuss (in Section~\ref{sec:abstractions}) these abstractions and the flexibility of the lattice structures in accommodating future graph data models and query languages. This will solidify the understanding of the fundamental principles of graph data extraction, exchange, processing, and analysis, as illustrated in Figure~\ref{fig:graphpipelines}.

A second important element is the vision of an ecosystem governing big-graph processing systems  and enabling the tuning of various components such as OLAP/OLTP operations, workloads, standards and performance needs  (Section~\ref{sec:ecosystems}). These aspects make the big processing systems more complicated than what seen in the last decade. Figure~\ref{fig:graphpipelines} provides a high-level perception of this complexity in terms of inputs, outputs, processing needs, and final consumption of graph data.

A third element is how to understand and control performance in these future ecosystems (Section~\ref{sec:performance}). We have important challenges to overcome in performance, from methodological aspects about performing meaningful, tractable, and reproducible experiments, to practical aspects regarding the trade-off of scalability with portability and interoperability.

\begin{figure}[!t]
\centering
\fcolorbox{blue}{blue!5}{
\begin{minipage}{5in}
{\bf Sidebar A: A Joint Effort by the Computer Systems and Data Management Communities}\\
\ \\
The authors of this article met in December 2019 in Dagstuhl for Seminar 19491 on Big Graph Processing Systems.\footnote{\url{ https://www.dagstuhl.de/en/program/calendar/semhp/?semnr=19491}} The seminar gathered a diverse group of 41 high-quality researchers from the data management and large-scale systems communities. It was an excellent opportunity to start the discussion about next-decade opportunities and challenges for graph processing.
\ \\
This is a community publication. The first four authors co-organized the community-event leading to this article and coordinated the creation of this manuscript. All other authors contributed equally to this research.

\end{minipage}%
}
\end{figure}

\section{Abstractions} \label{sec:abstractions}

Abstractions are widely used in programming languages, computational systems, database systems, etc., to conceal technical aspects in favor of more user-friendly, domain-oriented logical views. Currently, users have to choose from a large spectrum of \textbf{graph data models} that are similar, but differ in terms of \textbf{expressiveness}, cost, and intended use for \textbf{querying} and \textbf{analytics}. This ‘\textbf{abstraction soup}’ poses significant challenges to be solved for the future.

\begin{figure}[!t]
  \centering
  \includegraphics[width=0.75\linewidth]{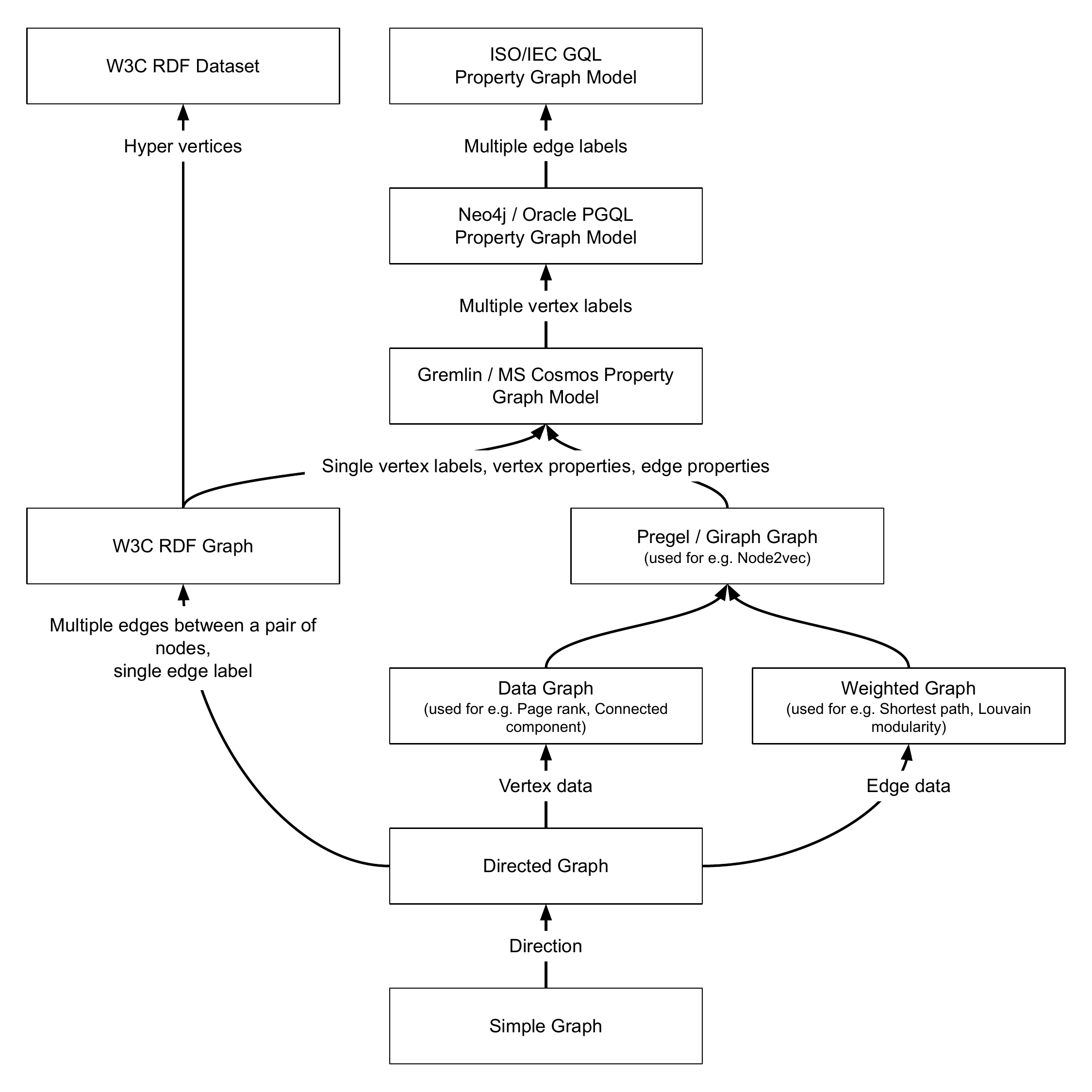}
  \caption{ Example lattice showing graph data model variants with their model characteristics, cf. [8].}
  \Description{A lattice of graph data models.}
  \label{fig:latticedatamodels} 
\end{figure}

\subsection{Understanding data models} \label{sec:abstractions:datamodels}

Today, graph data management faces many data models (directed graphs, RDF, variants of Property Graphs, etc.), with key challenges: (a) deciding which \textbf{data model} to choose per \textbf{use case}, and (b) mastering \textbf{interoperability} of data models where data from different models is combined (as in the left-hand side of Figure~\ref{fig:graphpipelines}). Both challenges require deepening our understanding of data models regarding: 

\begin{enumerate}

    \item How do \textbf{humans conceptualize} data and data operations? How do data models and their respective operators support or hinder the human thought-process? Can we measure how “natural” or “intuitive” data models and their operators are? 
    
    \item How can we quantify, compare, and (partially) order the (modelling and operational) \textbf{expressive power} of data models? Concretely, Figure~\ref{fig:latticedatamodels} illustrates a lattice for a selection of graph data-models. Read bottom-up, this lattice shows which characteristic has to be added to a graph data model to obtain a model of richer expressiveness. The figure also underlines the diversity of data models used in theory, algorithms, standards, and relevant\footnote{The figure does not aim to provide a complete list of Graph DBMS products. Please consult, e.g., \url{https://db-engines.com/en/ranking/graph+dbms} and other market surveys for comprehensive overviews.} industry systems. How to extend this comparative understanding across multiple data model families, such as graph, relational, document? Which costs and benefits of choosing one model over another?

    \item Interoperability between different data models can be achieved by means of \textbf{mappings} (semantic assertions across concepts in different data models) or with \textbf{direct translations} (e.g. W3C’s R2RML). Are there general ways or building blocks for expressing such mappings (e.g. category theory)?
    
\end{enumerate}
    
Studying (1) requires foremost investigating people working with data and data models, which is uncommon in the data management field and should be conducted collaboratively with other fields, such as human-computer interaction (HCI). Work on HCI and graphs exists, e.g., in HILDA workshops at Sigmod; however, these are not exploring the search space of graph data models. Studying (2) and (3) can build on existing work in database theory, but can also leverage findings from neighboring computer science communities on comparison, featurization, graph summarization, visualization and transformation of models. As an example, graph summarization [26] has been widely exploited in order to provide succinct representations of graph properties in graph mining [27] but they have seldom been used by graph processing systems to make processing more efficient, more effective, and more user-centered. For instance, approximate query processing for property graphs cannot rely on sampling as done by its relational counterpart and might need to use quotient summaries for query answering.

\subsection{Logic-based and declarative formalisms} \label{sec:abstractions:formalisms}

Logic provides a unifying formalism for \textit{expressing} queries, optimizations, integrity constraints, and integration rules. Starting from Codd’s seminal insight \textbf{relating logical formulae to relational queries} [3], many First Order (FO) logic fragments have been used to formally define query languages with desirable properties such as decidable evaluation. Graph query languages are essentially a syntactic variant of FO augmented with \textbf{recursive} capabilities. Logic provides a yardstick for \textit{reasoning} about graph queries and graph constraints. Indeed, a promising line of research is the application of formal tools, such as model checking, theorem proving [4], and testing, to establishing the \textit{functional correctness} of complex graph-processing systems, in general, and of graph database systems, in particular.

The influence of logic is pivotal not only to database languages, but also as a foundation for combining logical reasoning with statistical learning in AI. Logical reasoning derives categorical notions about a piece of data by logical deduction. Statistical learning derives categorical notions by learning statistical models on known data and applying it to new data. Both leverage the topological structure of graphs (ontologies and knowledge graphs,\footnote{A recent practical example it the COVID-19 Knowledge Graph: \url{https://covidgraph.org/}} graph embeddings (e.g. Node2vec)$^3$ to produce better insights than on non-connected data). However, both happen to be isolated. Combining both techniques can provide crucial advancements. As an example, deep learning (unsupervised feature learning) applied to graphs allows us to infer structural regularities and obtain meaningful representations for graphs that can be further leveraged by indexing and querying mechanisms in graph databases and exploited for logical reasoning. As another example, probabilistic models and causal relationships can be naturally encoded in property graphs and are the basis of advanced graph neural networks.\footnote{Zonghan Wu, Shirui Pan, Fengwen Chen, Guodong Long, Chengqi Zhang, Philip S. Yu (2019) A Comprehensive Survey on Graph Neural Networks. CoRR abs/1901.00596} Property graphs allow us to synthesize more accurate models for machine learning pipelines due to their inherent expressivity and embedded domain knowledge. These considerations unveil important open questions as follows: How can statistical learning and graph processing and reasoning be combined and integrated? Which underlying formalisms make this possible? How to weigh between the two mechanisms?

\subsection{Algebraic operators for graph processing} \label{sec:abstractions:algebraicoperators}

Currently, there is no standard graph algebra. The outcome of the \textbf{GQL} standardization project (cf. Section~\ref{sec:ecosystems:standards}) could influence the design of a graph algebra alongside existing and emerging use-cases [12]. However, next-generation graph processing systems should address questions about their algebraic components. What are fundamental operators of this \textbf{algebra}, compared to other algebras (relation algebra, group algebra, quiver or path algebra, incidence algebra or monadic algebra comprehensions)? What \textbf{core graph-algebra} should be supported by graph processing systems? Are there graph analytical operators to include in this algebra? Can this graph algebra be combined and integrated with an algebra of types, to make type-systems more expressive and to facilitate type checking?

A “relational-like” graph algebra able to express all the first-order queries [5] and enhanced with a graph pattern matching operator [28] seems like a good starting point. However, the most interesting graph-oriented queries are \textbf{navigational} (i.e., reachability queries) and cannot be expressed with limited recursion of relational algebra [7, 8]. Furthermore, relational algebra is a \textbf{closed} algebra, i.e., input(s) and output of each operator is a relation, which make relational algebra operators composable. Should we aim for a closed graph algebra that encompasses both relations and graphs? 

Current graph query engines combine algebra operators and ad-hoc graph algorithms into \textbf{complex workloads} (cf. Section~\ref{sec:ecosystems:workloads}). This complicates implementation and affects performance. An implementation based on a single algebra also seems utopic. A query language with general Turing Machine capabilities (like a programming language), however, entails tractability and feasibility problems [6]. Algebraic operators that work in both centralized and distributed environments, and that can be exploited by both graph algorithms and machine learning models, including GNNs, graphlets, and graph embeddings, could be highly desirable for the future.

\section{Ecosystems} \label{sec:ecosystems}

Ecosystems behave differently from mere systems of systems, because they couple many systems developed for different purposes and with different processes. Figure~\ref{fig:graphpipelines} exemplifies the complexity of a graph processing ecosystem through high-performance OLAP and OLTP pipelines working together. What are the ecosystem-related challenges?

\subsection{Workloads in graph processing ecosystems} \label{sec:ecosystems:workloads} \label{sec:ecosystems:wl}

Workloads affect both the functional requirements (what a graph processing ecosystem will be capable of doing) and the non-functional (how well). Survey data [12] point to pipelines as in Figure~\ref{fig:graphpipelines}: complex workflows, combining heterogeneous queries and algorithms, managing and processing diverse datasets, with characteristics summarized by Sidebar~B. 

In Figure~\ref{fig:graphpipelines}, graph processing links to domain-specific processing ecosystems, including simulation and numerical methods in science and engineering, aggregation and modeling in business analytics, and ranking and recommendation in social media; and to general processing, including machine learning.

\begin{figure}[!t]
\centering
\fcolorbox{green}{green!5}{
\begin{minipage}{5in}
{\bf Sidebar B: In-Depth: Known properties of graph-processing workloads}\\
\ \\
Graph workloads may exhibit several properties:
\begin{enumerate}

    \item Graph workloads are useful for many, vastly diverse domains [12-15]. Notable features include edge orientation, properties/timestamps for edges and nodes; graph methods (neighborhood statistics, pathfinding and traversal, and subgraph mining); programming models (think-like-a-vertex, think-like-an-edge, and think-like-a-subgraph); diverse graph sizes, including trillion-edge graphs [15]; and query and process selectivities [1].

    \item Graph workloads can be highly irregular, mixing (short-term) data-intensive and compute-intensive phases [15]. The source of irregularity, e.g., different datasets and algorithms, and computing platforms, affects performance greatly. Their interdependency forms the Hardware-Platform-Algorithm-Dataset (HPAD) Law [13]. 

    \item Graph processing uses a complex pipeline, combining a variety of tasks other than querying and algorithms [12, 27]. From traditional data-management, workloads include: transactional (OLTP) workloads in multi-user environments, with many short, discrete, likely atomic transactions; and analytical (OLAP) workloads with fewer users but complex and resource-intensive queries or processing jobs, with longer runtime (e.g., minutes). Popular tasks also include ETL; visualization;  cleaning; mining; and debugging and testing, including synthetic graph generation. 

    \item Scalability, interactivity, and usability affect how graph users construct their workloads [12].

\end{enumerate}

\end{minipage}%
}
\end{figure}

\subsection{Standards for data models and query languages} \label{sec:ecosystems:standards}

Graph-processing ecosystem standards can provide a common technical foundation, thereby increasing the mobility of applications, tooling, developers, users, and stakeholders. Standards for both OLTP and OLAP workloads should standardize the data model, the data manipulation and data definition language, and the exchange formats. They should be easily adoptable by existing implementations, and also enable new implementations in the SQL-based technological landscape. It is important that standards reflect existing industry practices, by following widely used graph query languages. To this end, in 2019, ISO/IEC started a project to define a new graph query language, GQL. GQL is backed by backing of ten national standards bodies with representatives from major vendors in the industry and supported by the property graph community as represented by the Linked Data Benchmarks Council (LDBC).\footnote{\url{http://ldbcouncil.org/}}

With an initial focus on transactional workloads, GQL will support composable graph querying over multiple possibly overlapping graphs using enhanced Regular Path Queries (RPQs) [7], graph transformation (views), and graph updating capabilities. GQL enhances RPQs with pattern quantification, ranking, and path-aggregation. Syntactically, GQL combines SQL-style with visual graph patterns pioneered by Cypher [14].

Long-term, worthwhile for standardization are building blocks of graph algorithms, analytical APIs and workflow definitions, graph embedding techniques, and benchmarks (cf. [21]). However, broad adoption for these aspects requires maturation.

\begin{figure}[!t]
  \centering
  \includegraphics[width=\linewidth]{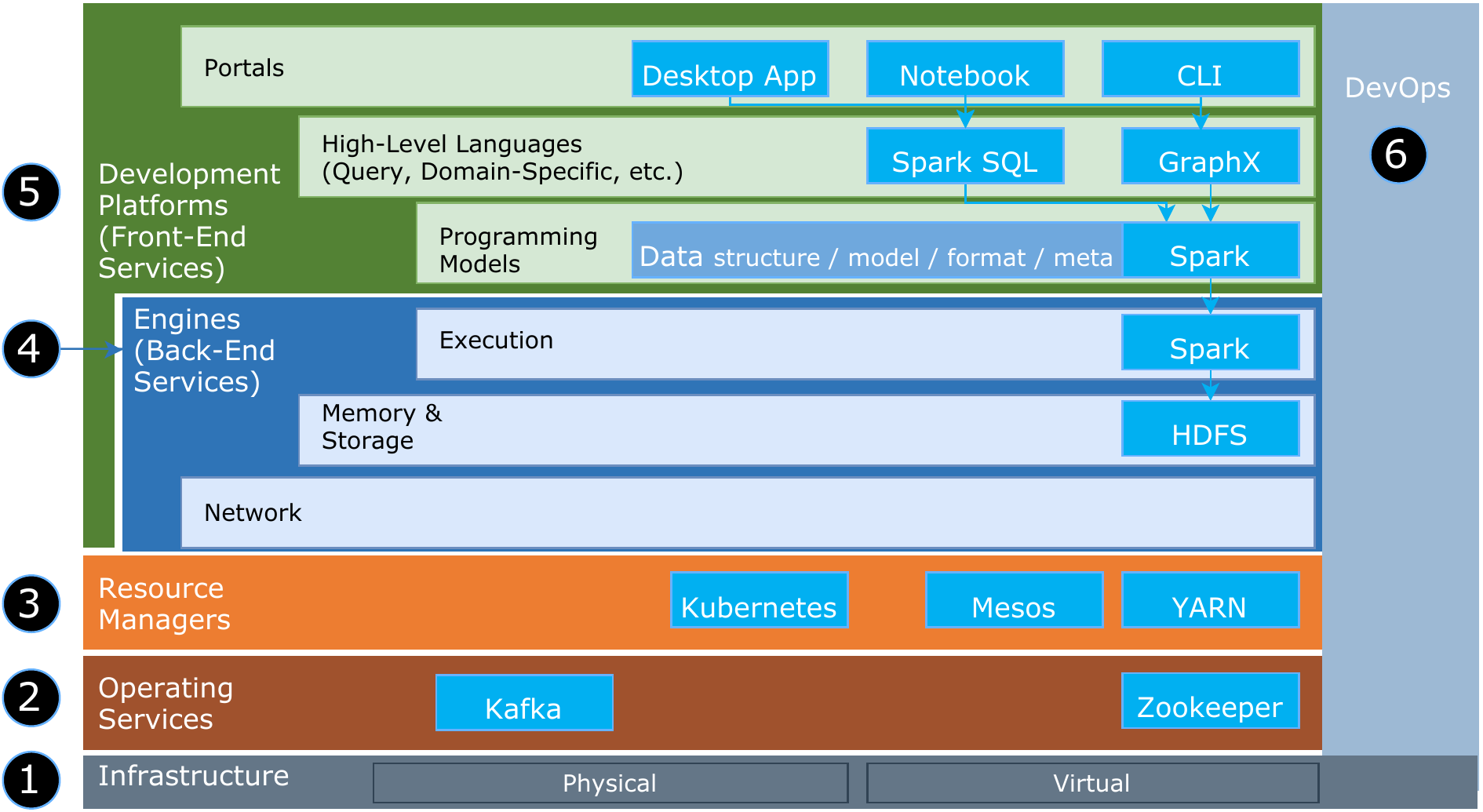}
  \caption{A reference architecture for graph processing ecosystems. The infrastructure layer (Layer 1) provides physical and virtual resources. The operating services layer (L2) provides services across resources, including data streaming and synchronization. The resource managers layer (L3) provides static and dynamic resource management and scheduling across resources. The back-end (L4) and front-end (L5) layers represent specialization efforts. Conversely, layers L2 and L3 may generalize techniques initially developed in L4-5.}
  \Description{A reference architecture for graph processing ecosystems.}
  \label{fig:refarchi} 
\end{figure}

\subsection{Reference architecture} \label{sec:ecosystems:referencearchitecture} \label{sec:ecosystems:refarchi}

We identify the challenge of defining a reference architecture for big graph processing. The early definition of a reference architecture has greatly benefitted the discussion around the design, development, and deployment of cloud and grid computing solutions [17]. 

For big graph processing, our main insight is that many graph processing ecosystems match the \textbf{common reference architecture of datacenters} [18], from which Figure~\ref{fig:refarchi} derives. The Spark ecosystem depicted here is one among thousands of possible instantiations. The challenge is to capture the evolving graph-processing field.

\subsection{Beyond scale-up vs. scale-out} \label{sec:ecosystems:scaling}

Many graph platforms  focus either on scale-up or on scale-out. Each has relative advantages [15]. Beyond merely reconciling scale-up and scale-out, we envision a \textbf{scalability continuum}: given a diverse workload, the ecosystem would decide automatically how to run it, and on what kind of heterogeneous infrastructure, meeting Service Level Agreements. 

There are numerous mechanisms and techniques to enforce scale-up and scale-out decisions, such as data and work partitioning; and migration, offloading, replication, and elastic scaling. All decisions can be taken statically or dynamically, using various optimization and learning techniques.

\subsection{Dynamic and streaming aspects} \label{sec:ecosystems:dynamic} \label{sec:ecosystems:streaming}

Future graph processing ecosystems should cope with dynamic and streaming graph data. A \textbf{dynamic graph} extends the standard notion of a graph to account for updates (e.g., insertions, changes, deletions) such that the current and previous states can be seamlessly queried. \textbf{Streaming graphs} can grow indefinitely, as new data arrives. They are typically unbounded thus the underlying systems are unable to keep the entire graph state. The sliding window semantics [9] allows to unify the two notions, with insertions and deletions being considered as arrivals and removals from the window.  

Since current streaming processing technologies are fairly simple, e.g. aggregations and projections as in industrial graph processing libraries (e.g. Gelly on Apache Flink), the need for “complex graph data streams” is evident along with more advanced graph analytics and ML ad-hoc operators. 
Another research challenge is to \textbf{identify the graph query processing operators} that can be evaluated on dynamic and streaming graphs while taking into account recursive operators [10, 11] and path-oriented semantics, as needed for standard query languages such as GQL and G-Core [19]. 

The graph processing platforms are also dynamic: Discovering, understanding, and controlling the \textbf{dynamic phenomena} that occur in complex graph processing ecosystems is an open challenge. As graph processing ecosystems become more mainstream and are embedded in larger data-processing pipelines, we expect to increasingly observe known systems phenomena such as performance variability, the presence of cascading failures, and autoscaling resources. What new phenomena will emerge? What programming abstractions [16] and systems techniques can respond to them?

\section{Performance} \label{sec:performance}

Graph processing raises unique performance challenges, from the lack of a widely used performance metric besides response time, to the methodological problem of comparing graph processing systems across architectures and tuning processes, to performance portability and reproducibility. Such challenges become even more daunting for graph processing ecosystems.

\subsection{Benchmarks, performance measurement, methodological aspects} \label{sec:performance:benchmarks} \label{sec:performance:measurement} \label{sec:performance:methodology}

Graph processing suffers from methodological issues as other computing disciplines [23,2]. Running comprehensive graph-processing experiments, especially at scale, lacks tractability [1], that is, ability to implement, deploy, and experiment within a reasonable amount of time and cost. As in other computing disciplines [23,2], we need new, reproducible experimental methodologies. 

Graph processing also raises unique challenges in performance measurement and benchmarking, related to complex workloads (see Sidebar~B) and data-pipelines (see Figure~\ref{fig:graphpipelines}). Even seemingly minute HPAD variations, e.g., the graph’s degree distribution, can have significant performance implications [24, 25]. The lack of interoperability (see Section~\ref{sec:performance:specialization}) hinders fair comparisons and benchmarking. Indexing and sampling techniques might prove to be useful to improve and predict the runtime and performance of graph queries [8, 29, 30], challenging the communities of large-scale systems, data management, data mining, and machine learning.  

Graph processing systems rely on complex runtimes combining software and hardware platforms. Capturing system-under-test performance, including parallelism, distribution, streaming vs. batch operation, etc., and testing the operation of possibly hundreds of libraries, services, and runtime systems present in real-world deployments, is daunting.

We envision a combination of approaches: As in other computing disciplines [23,2], we need new, reproducible experimental methodologies. Concrete questions arise: How to facilitate quick yet meaningful performance-testing? How to define more faithful metrics for executing a graph algorithm, query, program, or workflow? How to generate workloads with combined operations, covering temporal, spatial, and streaming aspects? How to benchmark pipelines including machine learning and simulation? We also need organizations such as the Linked Data Benchmark Council (LDBC) to curate benchmark sharing and to audit their use in practice.

\subsection{Specialization vs. portability and interoperability} \label{sec:performance:portability} \label{sec:performance:specialization} \label{sec:performance:interoperability}

There is considerable tension between specializing graph processing stacks for performance reasons and enabling productivity for the domain scientist, through portability and interoperability. 

\textbf{Specialization}, through custom software and especially hardware acceleration, leads to significant performance improvements. Specialization to graph-workloads (see Sidebar~B) focuses on diversity and irregularity\footnote{Irregularity could be seen as the opposite of the locality principle commonly leveraged in computing.} in graph processing: sheer dataset-scale (addressed by Pregel, later by the open-source project Giraph), the (truncated) power-law-like distributions for vertex degrees (PowerGraph), localized and community-oriented updates (GraphChi), diverse vertex-degree distributions across datasets (PGX.D, PowerLyra), irregular or non-local vertex access (Mosaic), affinity to specialized hardware (the BGL family, HAGGLE, rapids.ai), etc. 

The HPC domain proposed specialized abstractions and C++ libraries for them, and high-performance and efficient runtimes across heterogeneous hardware. Examples include BGL [21], CombBLAS, and GraphBLAS. Data management approaches, e.g., Neo4j, GEMS [22], and Cray’s Urika, focus on convenient query languages such as SPARQL and Cypher to ensure portability. Ongoing work also focuses on (custom) accelerators.

\textbf{Portability} through reusable components seems promising, but currently there exists no standard graph library or query language. Over 100 big-graph processing systems exist, but they do not support portability and soon graph-systems will need to support constantly evolving processes.  

Lastly, \textbf{interoperability} means integrating graph processing into broader workflows, with multi-domain tools. Integration with machine learning and data mining processes, and with simulation and decision making instruments, seems vital but is not supported by existing frameworks.

\subsection{A memex for big graph processing systems} \label{sec:performance:memex}

Inspired by Vannevar Bush’s 1940s concept of personal memex, and by a 2010s specialization into a Distributed Systems Memex [20], we posit it would be both interesting and useful to create a Big Graph Memex for collecting, archiving, and retrieving meaningful operational information about such systems. This could be beneficial for learning about and eradicating performance and related issues, for enabling more creative designs and extending automation, for meaningful and reproducible testing, as feedback building-block in smart graph processing, etc.

\section{Conclusion} \label{sec:conclusion}

Graphs are a mainstay abstraction in today’s data processing pipelines. How can future big graph processing and database systems  provide highly scalable, efficient and diversified querying and analytical capabilities, as demanded by real-world requirements? 

To tackle this question, we have undertaken a community approach. We started through a Dagstuhl Seminar and, shortly after, shaped the structured connections presented here. We have focused in this article on three interrelated elements: abstractions, ecosystems, and performance. For each of these elements, and across them, we have provided a view at what’s next.

Only time can tell if our predictions provided worthwhile directions to the community. In the meantime, join us in solving the problems of big graph processing. The future is \textit{big} graphs!

\section*{References}
\begingroup
\renewcommand\labelenumi{$[$\theenumi$]$}
\begin{enumerate}
    \item  Bonifati, A. et al. Graph Generators: State of the Art and Open Challenges. ACM Computing Surveys. April (2020). Online: \url{https://doi.org/10.1145/3379445}
    \item  Angriman, E. et al. (2019) Guidelines for Experimental Algorithmics: A Case Study in Network Analysis. Algorithms 2019, 12(7), 127.
    \item  Codd, E. F. A Relational Model of Data for Large Shared Data Banks”. Commun. ACM 13(6): (1970), 377-387.
    \item  Gonthier et al. A Machine-Checked Proof of the Odd Order Theorem. ITP (2013), 163-179.
    \item  Chandra A. K.. Theory of Database Queries. In Proc. Symposium on Principles of Database Systems, pages 1–9. ACM Press, (1988).
    \item  Aho, A. V.  and Ullman, J. D. Universality of data retrieval languages. In Proceedings of the 6th ACM SIGACT- SIGPLAN symposium on Principles of programming languages, pages 110–119. ACM Press, (1979).
    \item  Angles, R. et al. Foundations of modern query languages for graph databases. ACM Computing Surveys (CSUR), 50(5), Sept. (2017).
    \item  Bonifati, A. et al. Querying Graphs. Synthesis Lectures on Data Management. Morgan \& Claypool Publishers, (2018).
    \item  Babcock, B. et al. Models and Issues in Data Stream Systems. PODS (2002), 1-16.
    \item  Pacaci, A., Bonifati, A., and \"{O}zsu, M. T. Regular Path Query Evaluation on Streaming Graphs. ACM SIGMOD (2020). 
    \item  Bonifati, A., Dumbrava, S., Gallego Arias, E. J.: Certified Graph View Maintenance with Regular Datalog. Theory Pract. Log. Program. 18(3-4): 372-389 (2018).
    \item  Sahu, S. et al. The ubiquity of large graphs and surprising challenges of graph processing: extended survey. VLDB J. 29(2) (2020), 595-618.
    \item  Uta, A. et al., Exploring HPC and big data convergence: A graph processing study on Intel Knights Landing, in CLUSTER (2018).
    \item  Francis, N. et al.: Cypher: An Evolving Query Language for Property Graphs, SIGMOD (2019).
    \item  Salihoglu, S., Özsu, M. T.: Response to "Scale Up or Scale Out for Graph Processing". IEEE Internet Computing 22(5): 18-24 (2018).
    \item  Vasiliki, K., Vlassov, V. and Haridi, S.: "High-level programming abstractions for distributed graph processing." IEEE Transactions on Knowledge and Data Engineering 30.2 (2017), 305-324.
    \item  Foster, I. and Kesselman, C.: The Grid 2: Blueprint for a new computing infrastructure.    Elsevier, 2003.
    \item  Iosup, A. et al.: Massivizing Computer Systems: A Vision to Understand, Design, and Engineer Computer Ecosystems Through and Beyond Modern Distributed Systems. ICDCS (2018), 1224-1237.
    \item  Angles, R. et al.: G-CORE: A Core for Future Graph Query Languages. SIGMOD (2018), 1421-1432.
    \item  Iosup, A. et al.: The AtLarge Vision on the Design of Distributed Systems and
    Ecosystems. ICDCS (2019), 1765-1776. 
    \item  Siek, J. G., Lee, L.-Q. and Lumsdaine, A.: The Boost Graph Library: User Guide and
    Reference Manual. Addison-Wesley, (2002).
    \item  Castellana, V. G. et al.: In-Memory Graph Databases for Web-Scale Data. IEEE Computer 48(3), (2015), 24-35. 
    \item  Papadopoulos, A. V. et al.: Methodological Principles for Reproducible Performance Evaluation in Cloud Computing. IEEE Trans. on Sw. Eng. (2019). 
    \item  Iosup, A. et al.: LDBC Graphalytics: A Benchmark for Large-Scale Graph Analysis
    on Parallel and Distributed Platforms. PVLDB 9(13), (2016), 1317-1328.
    \item  Saleem, M. et al.: How Representative Is a SPARQL Benchmark? An Analysis of RDF Triplestore Benchmarks. WWW (2019), 1623-1633.
    \item  Liu, Y., Safavi, T., Dighe, A., Koutra, D.: Graph Summarization Methods and Applications: A Survey. ACM Comput. Surv. 51(3): 62:1-62:34 (2018).
    \item  Aggarwal, C-C., Wang, H.: Managing and Mining Graph Data. Advances in Database Systems 40, Springer 2010, ISBN 978-1-4419-6044-3.
    \item  He, H., Singh, A-K.: Graphs-at-a-time: query language and access methods for graph databases. SIGMOD Conference 2008: 405-418.
    \item  Leskovec, J., Faloutsos, C.: Sampling from large graphs. KDD 2006: 631-636.
    \item  Zhao, P., Han, J.: On Graph Query Optimization in Large Networks. Proc. VLDB Endow. 3(1): 340-351 (2010).
\end{enumerate}
\endgroup

\end{document}